\documentclass[superscriptaddress,aps,floats,onecolumn,twoside,floatfix]{revtex4}
\usepackage[english]{babel}

\usepackage{epsfig}
\usepackage{amsmath}
\usepackage{amssymb}
\usepackage{dsfont}
\usepackage{hyperref}
\usepackage{graphicx}
\usepackage{comment}
\usepackage{color}

\newcommand{\avg}[1]{\langle #1 \rangle}

\def\be{\begin{equation}}
\def\ee{\end{equation}}
\def\bea{\begin{eqnarray}}
\def\eea{\end{eqnarray}}
\def\dl{\langle}
\def\dr{\rangle}

\newcommand{\y}{z}

\begin{document}

\title{Aggregation models on hypergraphs}
\author{Diego Alberici}
\author{Pierluigi Contucci}
\author{Emanuele Mingione}
\author{Marco Molari\footnote{Present address: Laboratoire de Physique Statistique, Ecole Normale Superieure, 24 rue Lhomond, 75005 Paris, France}\begingroup
  \renewcommand\thefootnote{}\footnote{
  \textbf{Email adresses}: diego.alberici2@unibo.it (Diego Alberici) pierluigi.contucci@unibo.it (Pierluigi Contucci) emanuele.mingione2@unibo.it (Emanuele Mingione) marco.molari3@studio.unibo.it (Marco Molari - corresponding author)
  }%
  \addtocounter{footnote}{-1}%
  \endgroup}
\affiliation{Dipartimento di Matematica, University of Bologna, Piazza di Porta San Donato 5, 40126 Bologna, Italy }

\begin{abstract}
Following a newly introduced approach by Rasetti and Merelli we investigate the possibility to extract topological
information about the space where interacting systems are modelled. From the statistical datum of their observable
quantities, like the correlation functions, we show how to reconstruct the activities of their constitutive parts which embed
the topological information. The procedure is implemented on a class of polymer models on hypergraphs with hard-core interactions.
We show that the model fulfils a set of iterative relations for the partition function that generalise those introduced
by Heilmann and Lieb for the monomer-dimer case. After translating those relations into structural identities
for the correlation functions we use them to test the precision and the robustness of the inverse problem. Finally the
possible presence of a further interaction of peer-to-peer type is considered and a criterion to discover it is identified.\\

\end{abstract}

\maketitle

\section{Introduction and Results}

In a recent paper \cite{rasetti2015topological} a new perspective for the general problem of data analysis,
in the context of Big Data and Complex Systems, has been advanced. By probing the data space encoded as a set of
correlation functions, the information content of a phenomenological setting is embedded into a
\emph{field theory of data} based on an underlying topological space. This idea is deeply rooted into concepts
that have originated from theoretical physics. General Relativity, to mention one of the examples, is the gravitational
field theory that describes the motion of particles through space-time where their dynamics is fully determined
by the underlying curvature.

We propose here a very simplified realisation of that program that capitalises on the equivalence of field theories
with classical statistical mechanics \cite{feynman1948space,schwinger1958euclidean,symanzik1966euclidean} with the purpose of testing it using the inverse problem
approach. The models we consider are hard-core interacting polymer systems on high-dimensional networks (hypergraphs). 
The choice of this class of models is due to the diversity and richness of the phenomena they describe that span from
Physics \cite{chang1939statistical}, Biology \cite{o1990thermodynamic}, Computer Science \cite{bordenave2013matchings, karp1981maximum, zdeborova2006number}, and Social Sciences \cite{barra2013integration}.
We have in mind, in particular, applications in the the socio-technical setting of novel communication systems where
groups of people are present in chambers like those of the messaging systems, voip conference calls etc. From a
mathematical point of view those are aggregation models of particles that cannot occupy at the same time more than
one state (hard-core constraint): in the specific example of the messaging systems an individual is either silent, the
monomer state, in a two body conversation, the dimer state, in a three body conversation state called trimer and so on. While the
old style phone calls were well described by a standard monomer-dimer model the novel technologies allow for the
contemporary presence of multiple individuals in the same virtual room thus requiring higher order objects like
hypergraphs for the underlying space and polymers for the fields that represent their state.

In our model the configurations of the system are determined by the occupation number on the elements of the
hypergraph (vertices, edges and faces) that takes only two values $0$ and $1$. We limit the analysis to the
rank three case (conversation with maximum three bodies in the mentioned example) but the generalisation to higher ranks is straightforward. The model is assigned by a set of positive weights,
the activities, associated to each hyperedge. These weights describe the strength of connections and identify the topology of the hypergraph trough, for instance, the persistent topology methods developed in \cite{frosini1992measuring, verri1993use}, in \cite{edelsbrunner2002topological,edelsbrunner2008persistent} and used in \cite{petri2014homological}. A threshold for the activities could be decided, and the hyperedges below this threshold deleted from the original hypergraph. Instead of studying the topology at an arbitrary threshold, the persistent topology approach consists in exploring the whole filtration of hypergraphs obtained by varying the threshold. Quoting \cite{rasetti2015topological}, ``this filtration process identifies those topological features which persist over a significant parameter range, qualifying them as candidates to be considered as signal, while those that have short-lived features can be assumed to characterize noise''. Afterwards this topological signal can be used to compare and classify different datasets.

Our first result is of rigorous mathematical nature: the identification of an iterative relation for the partition function of the model which generalises
the Heilmann-Lieb identity \cite{heilmann1972theory}. While this relation is introduced in a hypergraph theoretical setting we show
that it implies a set of identities directly expressible in terms of the correlation functions of the
associated probability measure. They act as a constitutive family of equations for the model that we use in our test
and turn out to be an essential tool toward an efficient control of the inverse problem, i.e. the basic question: from a (full or partial)
set of the correlation functions can we recover the value of the activities for all the hyperedges?

This work provides a positive answer to the previous question together with the possible limitations and contains
two conceptually different numerical methods which can be used to extract activities from the experimental correlations. 
The first inversion method is based on the maximisation of the \textit{likelihood} function
and works through a recursive gradient-descent algorithm partially inspired by the one used for the learning process in Boltzmann Machines \cite{ackley1985learning}. We tested its performance and found that it converges exponentially at a speed that does not
depend on the size of the hypergraph but is influenced by the magnitude of the activities. In particular the convergence speed decreases at higher values of the activities, as expected when reaching the full packing regime. The second method is based on the maximisation of the \textit{pseudo-likelihood} function when additional experimental correlations are known. This has the advantage that it can
be applied in a much simpler manner since it provides an explicit expression for the activities. 

Finally we study the effects of the presence of a further interaction acting among monomers in the hypergraph. 
In socio-technical systems this kind of interaction generated by peer-to-peer effects is often very relevant. 
The extra structure that comes with it is codified by another hypergraph built on the same set of vertices which, in general, is 
different and independent from the previous one. The two networks indeed can be seen as a bilayer structure 
like those analysed in \cite{GinestraMulti}. We concentrated on the problem of probing the presence of such an interaction from the set of experimental correlations, and found that the comparison between the two previously introduced inversion methods provides a good test for the detection of the interaction. Moreover, in the high interaction limit, we show how the same comparison can also be used to numerically estimate the parameter magnitude.

\section{The theoretical framework}\label{sec:model}
Let $H=V\cup K$ be a \textit{hypergraph} of rank 3, that is a set of vertices $V$ and hyperedges $K$ where  $K=E\cup F$ is an union of edges $E$ and faces $F$ (our notation naturally generalises to arbitrary rank).
On this topological space we consider configurations of \textit{polymers}, precisely monomers (single particles occupying a vertex), dimers (2-particles occupying an edge), trimers (3-particles occupying a face). Polymers display mutual hard-core interaction: no region of the space can be touched by more than one polymer. At the same time we require all the vertices of the hypergraph to be covered by either a monomer or one of the vertices of a polymer. This last condition that we call \textit{filling}, fully specifies the ensemble and should not be confused with the \textit{full-packing} one where monomers are not allowed. 

A suitable way to represent the allowed configurations is to introduce the \textit{occupancy variables}
$\alpha=\big(\alpha_h\big)_{h\in H}\in \{0,1\}^{H}$
with the \textit{hard-core filling} condition
\begin{equation}\label{hardcore}
\alpha_v+\sum_{\substack{e\in E:\\ e \ni v}}\alpha_e+ \sum_{\substack{f\in F:\\ f \ni v}}\alpha_f=1,\;\; v\in V.
\end{equation}
Notice that because of \eqref{hardcore}, for any vertex $v\in V$ the quantity $\alpha_v$, that represents the monomer occupancy of the vertex $v$, can always be expressed as a function of the dimer and trimer occupancy variables.
It is convenient to introduce the admissibility characteristic function $C:\{0,1\}^{H}\to \{0,1\}$ defined as
\begin{equation}\label{hardcorefun}
C(\alpha) = \begin{cases}
       1 & \text{if \eqref{hardcore} holds}\\
       0 & \text{otherwise}
       \end{cases}\,.
\end{equation}
To fully specify the model we introduce the {\it polymer activity} of each hyperedge, that is a positive number that measures the propensity of the hyperedge to be occupied by a corresponding polymer. One can show with an elementary computation that
the vertex activities can be reabsorbed into the remaining parameters or factorised out of the partition function. We denote by $\y_e,\ e\in E$  the {\it edge activities} (or \textit{dimer activities}) and by $z_f,\ f\in F$  the {\it face activities} (or \textit{trimer activities}).
The topological and analytical data, namely $H$ and $z$, fully determine a \textit{probability measure} associated to configurations:
\begin{equation}\label{gibbsmeasure}
\mu_z(\alpha) \; = \; \frac{C(\alpha)\prod_{e\in { E}}\y_e^{\alpha_e}\prod_{f\in { F}}z_f^{\alpha_f}}{Z(z)} \;,\quad\alpha\in\{0,1\}^H
\end{equation}
where $Z$ is the normalisation factor usually called \textit{partition function}.
We denote by $\langle\,\cdot\,\rangle$ the average w.r.t. the probability measure \eqref{gibbsmeasure}.\\
\vskip 0.2cm

Defining $E(v)$ the set of edges with one vertex in $v$ and $F(v)$ the set of faces with one vertex in $v$, one can prove that the following iterative relation holds:
\begin{equation}\label{recursionHL}
Z_{H} \; = \; Z_{H- v} + \sum_{e\in E(v)}\y_e\, Z_{H- e}
+ \sum_{f\in F(v)}z_f\, Z_{H- f} \;
\end{equation}
which generalises the Heilmann-Lieb relation for monomer-dimer systems \cite{heilmann1972theory,heilmann1970monomers}.
In eq.\eqref{recursionHL}, $H- v$ denotes the hypergraph where the vertex $v$ has been removed together with the hyperedges in $E(v)\cup F(v)$; $H- e$ stands for $H- v_1- v_2$ where $e=\{v_1,v_2\}$; $H- f$ stands for $H- v_1- v_2- v_3$ where $f=\{v_1,v_2,v_3\}$.\\

The previous family of relations \eqref{recursionHL} for the partition function of the model implies the following \textit{topological constraint relations} for the correlation functions.
For every edge $e=\{i,j\}$ and for every observable $g$ that does not depend on $\alpha_e$, $\alpha_i$ and $\alpha_j$ it holds:
\be
 \langle \alpha_e\, g \rangle \,=\, \y_e\, \langle  \alpha_{i}\alpha_{j}\, g \rangle \;.
\ee
Similarly, for every face $f=\{i,j,l\}$ and for every observable $g$ that does not depend on $\alpha_f$ , $\alpha_{i}$, $\alpha_{j}$
and $\alpha_{l}$ it holds:
\be
\langle \alpha_f\, g \rangle \,=\, z_f\, \langle  \alpha_{i}\alpha_{j}\alpha_{l}\, g \rangle \;.
\ee
In particular for $g\equiv 1$ one obtains an explicit expression of the activities in terms of correlations
\begin{equation}\label{hcrelations}
 \y_e \,=\, \frac{ \langle \alpha_e\rangle} {\langle \alpha_i\alpha_j \rangle} \quad,\quad z_f \,=\, \frac{\langle \alpha_f\rangle} {\langle\alpha_i\alpha_j \alpha_l\rangle} \;.
\end{equation}

\section{The inverse problem} \label{sec:inverse}
In the last few years several new ideas and techniques have been developed \cite{aurell2012inverse, sessak2009small, roudi2009ising} for the \textit{inverse problem} of the Ising model. We will discuss the inverse problem for the class of hard-core polymer models introduced in the previous section.
The general task is to extract the parameters of a given theoretical model from experimental measures on the observables.
The problem clearly displays different features according to the types of data that become available. In this work we will focus on two experimental database settings. In the first one the dataset is composed by the empirical densities of dimers and trimers, while in the second one some empirical correlations for the monomers are also included:
\begin{itemize}
\item[A)] the \textit{empirical polymer densities}, that is $\dl \alpha_e \dr_{\mathrm{exp}}$ for very edge $e\in E$
and $\dl \alpha_f \dr_{\mathrm{exp}}$ for every face $f\in F$;
\item[B)] the previous \textit{empirical polymer densities} plus the \textit{empirical monomer correlations}, that is
$\dl\alpha_i\alpha_j\dr_{\mathrm{exp}}$ for every edge $e=\{i,j\}\in E$ and $\dl\alpha_i\alpha_j\alpha_l\dr_{\mathrm{exp}}$ for every face $f=\{i,j,l\}\in F$.
\end{itemize}
The symbol $\dl\;\dr_{\mathrm{exp}}$ denotes the empirical average, that is if $M$ polymer configurations $\alpha^{(1)},\dots,\alpha^{(s)}$ are observed independently then $\dl g \dr_{\mathrm{exp}} \equiv \frac{1}{M}\sum_{s=1}^M g(\alpha^{(s)})\,$.

\subsection{The Kullback-Leibler method} \label{sec:KL}
In case A) the \textit{Maximum Likelihood Estimation} (MLE) can be used. Let us denote by $\mu_{z}$ and  by $\dl \;\dr_{z}$ respectively the probability measure defined by \eqref{gibbsmeasure} and the associated expectation. It is possible to prove (see Appendix) that the MLE of the polymer activities $z^*=(z_k^*)_{k\in K}$ satisfies the following set of $|K|$ conditions
\be\label{mlestimation}
\begin{split}
\dl \alpha_e\dr_{z^*} \,&=\, \dl \alpha_e \dr_{\mathrm{exp}}\,,\quad e\in E \\
\dl \alpha_f\dr_{z^*} \,&=\, \dl \alpha_f \dr_{\mathrm{exp}}\,,\quad f\in F \;.
\end{split}
\ee
The set of equations \eqref{mlestimation} determines implicitly the activities.
We approach its solution by means of a gradient descent algorithm since the Maximum Likelihood function is a concave function. 
Precisely at step $n+1$ ($n\geq0$) we update the vector of polymer activities $z^{(n)}\equiv\big(z^{(n)}_k\big)_{k\in K}$ as follows
\be \label{eq:BL}
z^{(n+1)} \,=\, z^{(n)} - \eta^{(n+1)}\, \frac {\nabla(z^{(n)})} {\sqrt{\sum_{k\in K} \big(\partial_{k}(z^{(n)})\big)^2}} \;.
\ee
The vector $\nabla(z) \equiv \big(\partial_{k}(z)\big)_{k\in K}$ is the gradient of the Kullback-Leibler divergence $D_{KL}(\mu_z|\mu^*)$, defined by:
\be \label{eq:BLgrad}
\partial_k(z) = - \frac{\dl\alpha_k\dr_{\mathrm{exp}}-\dl\alpha_k\dr_{\mathrm{z}}}{z_k}
\ee
and it gives to the update step $\Delta \y^{(n+1)} \equiv \y^{(n+1)} - \y^{(n)}$ the direction of the gradient of the likelihood function, or equivalently the direction of minus the Kullback-Leibler divergence gradient (see Appendix for the details).
The positive number $\eta^{(n+1)}$ tunes the magnitude of the update steps $\Delta \y^{(n+1)}$.  
By fixing $\eta^{(n)}\equiv \eta$, the speed of convergence of relation \eqref{eq:BL} is linear, while it can be improved by introducing an adaptive learning rate defined iteratively as:
\be \label{eq:eta}
\eta^{(n+1)} \,=\, \eta^{(n)}\, \exp \left\{ \gamma\, \frac{ \sum_{k\in K} \Delta z_k^{(n)}\, \Delta z_k^{(n-1)} } { \sqrt{\sum_{k\in K} \left(\Delta z_k^{(n)}\right)^2}\, \sqrt{\sum_{k\in K} \left(\Delta z_k^{(n-1)} \right)^2} } \right\}
\ee
$\gamma$ is a positive parameter to be chosen. The relation \eqref{eq:eta} is based on the scalar product between two consequent updates of the activities. If it is positive, which means that the last update steps $\Delta z_k^{(n)}$, $\Delta z_k^{(n-1)}$ were performed along similar directions, then the next update $\Delta z_k^{(n+1)}$ will have a greater magnitude. If it is negative, which means that the last two updates were performed along opposite directions, then we are in proximity of the solution and a greater precision is needed, so the magnitude of the next update step is diminished.

The recursion stops when the value of the activities $z^{(n_f)}$ is sufficiently close to the exact MLE solution of the inverse problem $z^{*}$. In our case we used two different stopping criteria. The first one can be used only when testing the performance of the algorithm on a priori known models, since it requires the knowledge of the exact values of the activities. In this case a value of precision $\epsilon_f > 0$ is chosen, and the recursion stops when the maximum relative error over the set of activities is less than $\epsilon_f$:
\be \label{crit1}
\epsilon^{(n_f)} = \max_{k \in K} \left| \frac{z^*_k - z_k^{(n_f)}}{z^*_k} \right| < \epsilon_f\;.
\ee
The second criterion can be applied when solving the inverse problem on experimental data, since it does not assume the knowledge of the exact value of the activities. Again a final precision value $\hat{\epsilon}_f > 0$ is chosen, and the recursion stops as soon as the set of equations \eqref{mlestimation} is satisfied with precision of at least $\hat{\epsilon}_f$:
\be \label{crit2}
 \hat\epsilon^{(n_f)} \,=\, \max_{k\in K}|\log\avg{\alpha_k}_{z^{(n_f)}}-\log\avg{\alpha_k}_{\exp}| < \hat{\epsilon}_f \; .
\ee

\textbf{Numerical tests}. In order to assess the reliability and stability of this method we performed numerical tests on the speed of convergence of the algorithm \eqref{eq:BL} to the solution of the equation \eqref{mlestimation} on random hypergraphs.

In particular we made use of a class of random hypergraph which represents the extension of the notion of Erd\H{o}s-R\'{e}ny random graph. This choice allows us to test the performance of our algorithm over different topologies. Moreover, real-world data is often constituted by many similar instances of the model, whose topologies can be considered as extracted from some random distribution
(see \cite{barra2013integration} for instance).

\begin{figure}
  \centering
  \includegraphics[width=\textwidth]{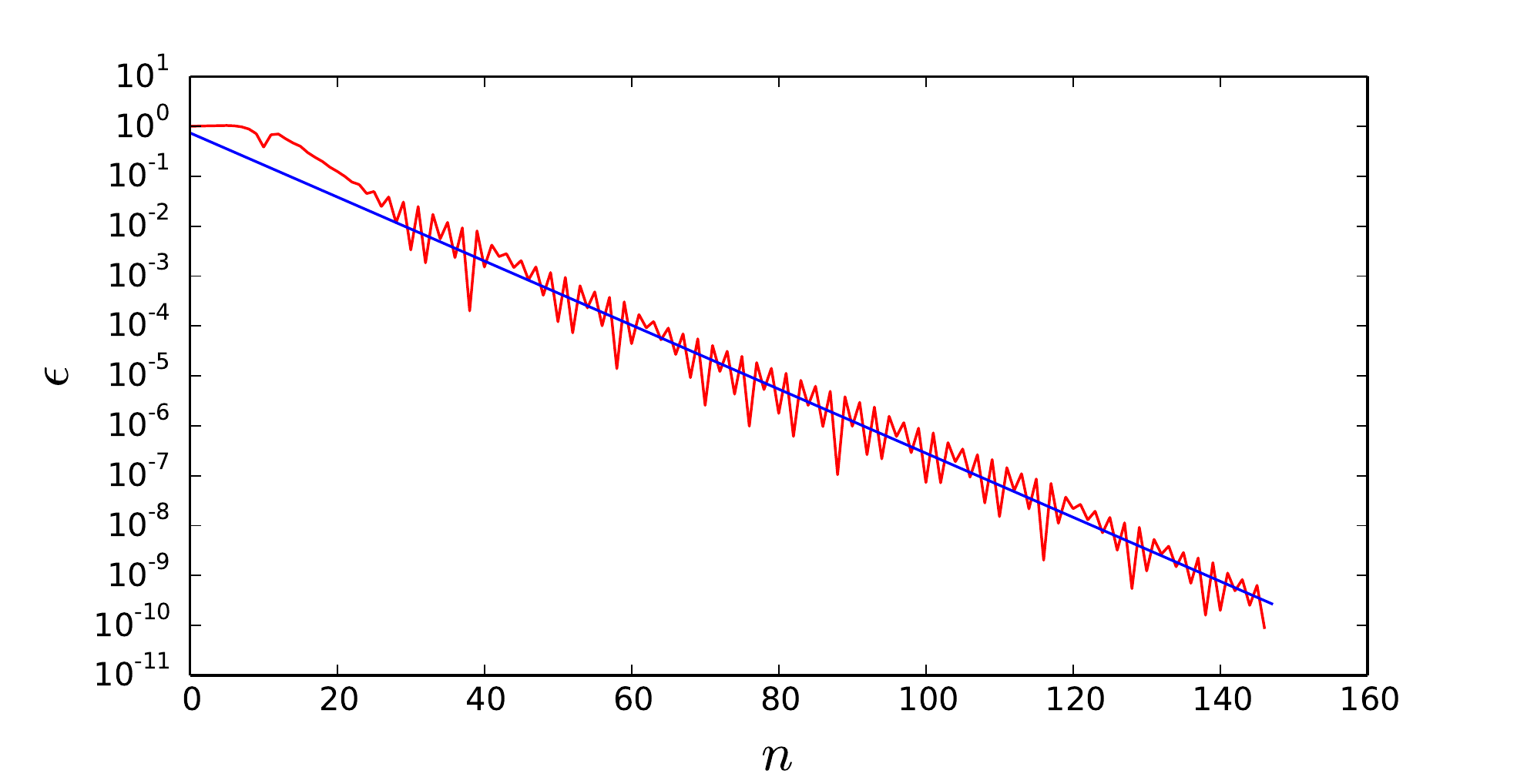}
  \caption{\textbf{Relative error of the gradient descent algorithm}: (Colour online) $\epsilon^{(n)} = |z^{(n)}_k - z_k|/ z_k$  versus number of iterations $n$ (red curve, linear-log scale). \textit{The convergence is exponentially fast in the number of iterations}: to test this hypothesis we performed a linear fit (blue line) according to the relation $\log \epsilon^{(n)} = -A\,n + B$.
We chose a random hypergraph with $N=15$, $c_1=c_2=1$ and $z_k=0.5$ for all $k\in K$. The fit is performed on the data after removing the initial $20\%$ of iterations.}
   \label{fig:learning_plot}
\end{figure}

\begin{figure}
  \includegraphics[width = \textwidth]{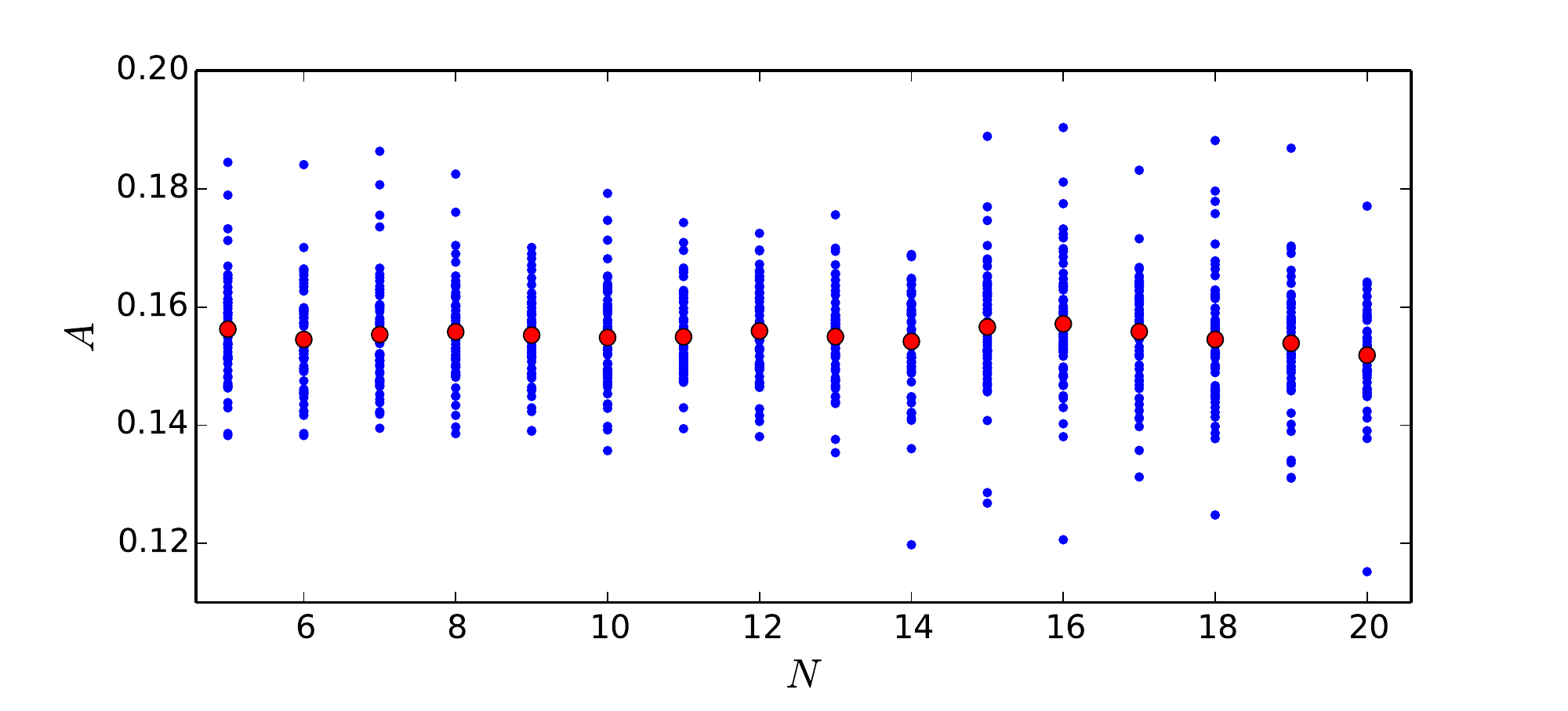}
  \caption{\textbf{Exponential rate of convergence $A$ of the gradient descent algorithm versus number of vertices $N$}, according to the fit $\log \epsilon^{(n)} = -A\,n + B$. (Colour online) The number of vertices ranges from $5$ to $20$. \textit{The distribution of the velocity of convergence does not seem to depend on the number of vertices.  Anyway we stress the fact that the larger $N$ is, the longer it takes to compute each step of the algorithm.}
 For each value of $N$ we performed $60$ trials on different random hypergraphs, taking always $c_1 = c_2 = 1$ and $z_k = 0.5$ for all $k\in K$.  The red dots represent the mean values of $A$ for each set of trials with the same value of $N$. To test the accuracy of the exponential fit we computed the correlation coefficient $R$: its average value and standard deviation over these $960$ tests are $R=-0.945\pm 0.023$.}
  \label{fig:fit_vs_N}
\end{figure}

\begin{figure}
  \centering
  \includegraphics[width = \textwidth]{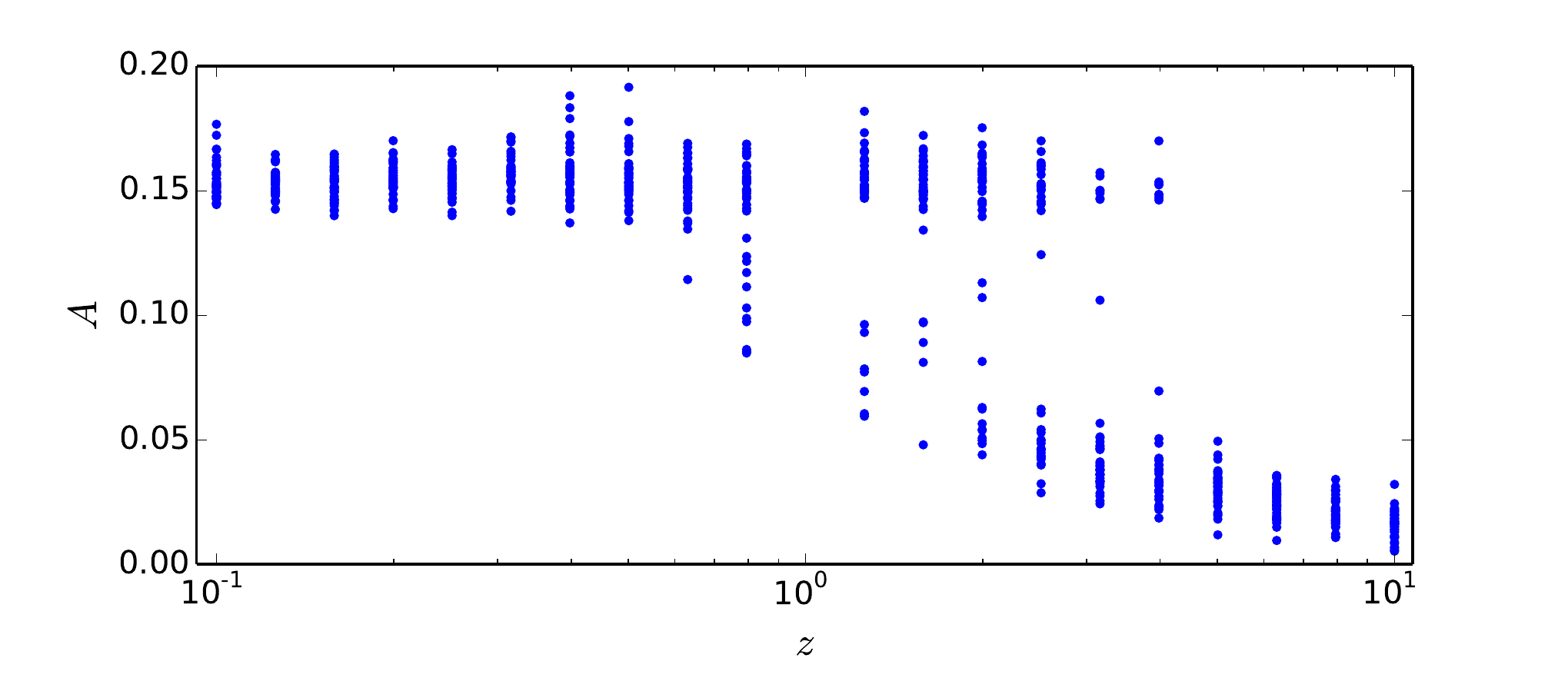}
  \caption{\textbf{Exponential rate of convergence $A$ of the gradient descent algorithm versus polymer activity $z$} (log-linear scale), according to the fit $\log \epsilon^{(n)} = -A\,n + B$. The activity is the same for each hyperedge ($z_k=z\;\forall\,k\in K$) and takes values $z=10^h$, $h=-1,-0.9,\dots,1$, excluding $h=0$ which is by default the starting point of our algorithm. \textit{The distribution of the rate of convergence depends on the intensity of the activity}: it is constant for $z\leq 10^{-0.2}$, then for $10^{-0.1}\leq z\leq 10^{0.6}$ it splits in two regions, and for $z\geq 10^{0.6}$ only the slower region survives and the rate of convergence decreases to zero.
For each value of $z$ we performed $40$ trials on different random hypergraphs, taking always $N = 20$, $c_1 = c_2 = 1$. The underlying hypothesis of exponential convergence is supported by the correlation coefficient $R = -0.968 \pm 0.028$ over these $800$ tests.}
  \label{fig:fit_vs_z_uniform}
\end{figure}

We observed that the convergence of the algorithm is exponentially fast in the number of iterations $n$ (Figure \ref{fig:learning_plot}). Moreover the distribution of the speed of convergence does not seem to depend on the number of vertices $N$ in the random hypergraph (Figure \ref{fig:fit_vs_N}). Anyway we stress the fact that the larger $N$ is, the longer it takes to compute each step of the algorithm, since the evaluation of $\avg{\alpha_k}_{z^{(n)}}$ is more demanding. On the contrary the speed of convergence depends on the intensity of the activities (Figure \ref{fig:fit_vs_z_uniform}). In particular in the limit of large polymer activity the exponential rate of convergence vanishes. This limit is equivalent to the \textit{full-packing} regime, in fact when polymer activities are high the presence of monomers is repressed in favour of higher order particles.

Precisely, to obtain these results, we have generated data as follows:
\begin{itemize}
  \item A random hypergraph $H = V\cup K$ over $N$ vertices is generated by placing each hyperedge independently. Each 2-edge is present with probability $2 c_1 / (N-1)$ and each 3-edge with probability $ 6 c_2 / (N-1) (N-2)$.
  \item An activity $z_k$ is assigned to each hyperedge $k \in K$. For simplicity when generating the dataset we chose $z_k = z$ constant for all $k \in K$. Details of this choice are specified in each case.
  \item All the possible monomer-dimer-trimer configurations $\alpha = (\alpha_k)_{k \in K}$ on the hypergraph are computed. We assign to each configuration its probability and we evaluate the expectations $\avg{\alpha_k}_{z}$.
\end{itemize}

The gradient descent algorithm was then applied, using as input parameters $\avg{\alpha_k}_{\exp} = \avg{\alpha_k}_{z}$. Clearly, this choice entails that $z$ solves eq. \eqref{mlestimation} and the recursion converges to the value $z^* = z$. We set $z^{(0)}_k=1$ for all $k \in K$ and $\gamma=0.2$. We used eq. \eqref{crit1} as stopping criterion setting $\epsilon_f = 10^{-10}$.

\subsection{The effects of an imitative perturbation}
It is important to notice that in case B) the number of observables is two times the number of degrees of freedom of the model defined by \eqref{gibbsmeasure}, since the dataset contains the \textit{empirical polymer densities} $\avg{\alpha_k}_{\exp}$ and the \textit{empirical monomer correlations} $\avg{\prod_{v\in k}\alpha_v}_{\exp}$ while the model is determined only by the activities $z_k$, $k\in K$.

A possible way to deal with this overdetermined case is to consider the \textit{Maximum Pseudo-Likelihood Estimation} (MPLE). This method can be seen as an approximation of the MLE where the joint distribution is replaced with a suitable conditional probability: we look at the probability to observe an occupied hyperedge conditionally on the states of all the others. It can be proven (see Appendix) that the MPLE of the activities $z^{**}$ satisfies the following set of $|K|$ conditions
\be\label{mplestimation}
\begin{split}
\dl \alpha_e \dr_{\mathrm{exp}} \,&=\, \y^{**}_e\, \dl\alpha_i\alpha_j\dr_{\mathrm{exp}}\,,\quad e=\{i,j\}\in E\\
\dl \alpha_f \dr_{\mathrm{exp}} \,&=\, z^{**}_f\, \dl\alpha_i\alpha_j\alpha_l\dr_{\mathrm{exp}}\,,\quad  f=\{i,j,l\}\in F \;.
\end{split}
\ee
We observe two important features: the analogy between \eqref{mplestimation} and the exact relations \eqref{hcrelations} and the fact that these relations provide
an explicit form for the activities.\\

Another way to exploit the additional information given by the empirical monomer correlations is to modify the model defined in \eqref{gibbsmeasure} by introducing a new family of parameters $J=(J_k)_{k\in K}$ that tune the monomer correlations:
\begin{equation}\label{gibbsmeasure2}
\mu_{z,J}(\alpha) \,=\, \frac{C(\alpha)\,
\prod_{k\in K}z^{\alpha_k}\, \exp\Big(\sum_{k\in K}J_k \prod_{v\in k}\alpha_v\Big)} {Z(z,J)} \,,\quad\alpha\in\{0,1\}^H \;.
\end{equation}
We denote by $\avg{\cdot}_{z,J}$ the average with respect to this probability measure. While this fact could appear as a mere technical device, it has instead a deep phenomenological meaning: the monomers can indeed directly interact beyond the hard-core repulsion, a situation largely expected in socio-technical systems due to the peer-to-peer effect among individuals.
In other words in the experiments the presence of a coupling $J$ between monomers cannot be excluded \textit{a priori}.
For this reason in this second part of our work we have generated the \textit{empirical polymer densities} and \textit{empirical monomer correlations} according to a perturbed distribution $\mu_{z,J}$.

The following extension of the Heilmann-Lieb identity for the partition function of the measure \eqref{gibbsmeasure2} holds:

\be \label{eq:liebJ}
Z_H \,=\, Z_{H-v}^* \,+\, \sum_{k\in K\atop k\ni v} z_k\, Z_{H-k}  \;, \quad v\in V
\ee
where in the partition function $Z_{H-v}^*$ a monomer activity $e^{J_{u\sim v}} := \prod_{k\in K,\,k\ni u,v} e^{J_k}$ is introduced on every vertex $u$ which was connected to $v$.
We call \textit{hypertree} a hypergraph $H$ such that, after having removed the edges included in some face, its line graph is a tree. On hypertrees the relation \eqref{eq:liebJ} provides the following useful estimate:
\be \label{eq:stima}
\dfrac{\avg{\alpha_k}_{z,J}}{\avg{\prod_{v\in k}\alpha_v}_{z,J}} \,=\, \dfrac{z_k}{\prod_{\substack{h\in K,\\ |h\cap k|>0}}e^{J_h}}\; \theta_k \;,\quad k\in K
\ee
where the term $\theta_k$ goes to $1$ as $z_p\,e^{-J_p}$ vanishes for every polymer $p\in K$ at distance $1$ from $k$, and even better:
\be
1 \,\leq\, \theta_k \,\leq\, \prod_{\substack{v\in V,\\ v\sim k}}\bigg(1+\sum_{\substack{p\in K,\\ p\ni v,\,|p\cap k|=0}}\,z_p\ \prod_{\substack{q\in K,\\ q\ni v,\,|q\cap k|=0}}e^{-J_q}\bigg) \;.
\ee

As said before, we have generated data $\avg{\alpha_k}_{\exp}$, $\avg{\prod_{v\in k}\alpha_v}_{\exp}$ according to the distribution \eqref{gibbsmeasure2} in the presence of an interaction $J \neq 0$: the quantities $\avg{\alpha_k}_{z,J}$ and $\avg{\prod_{v \in k} \alpha_v}_{z,J}$ have been computed exactly on random hypergraphs, following a procedure analogous to Section \ref{sec:KL}.
Starting from these data we have computed the MLE and MPLE as if the interaction was not present.
	We guessed that while the two resulting estimates $z^*$ and $z^{**}$ of the activities agree in case $J=0$, they may differ when $J \neq 0$, and thus they may be used to probe the presence of an interaction. To  make this guess more precise, we performed the following test, which could be applied also to real data.
\begin{itemize}
\item The gradient descent algorithm \eqref{eq:BL} is executed using as input $\avg{\alpha_k}_{\exp}=\avg{\alpha_k}_{z,J}$. If the algorithm converges, its limit is a vector of activities $z^*$ such that:
\be
 \avg{\alpha_k}_{z^*} \,=\, \avg{\alpha_k}_{z,J} \,,\quad k\in K\;.
\ee
We set $z^{(0)}_k=1$ and $\gamma=0.2$. We used eq. \eqref{crit2} as stopping criterion setting $\hat\epsilon_f = 10^{-5}$, together with a bound for the number of iterations that stops the recursion at $n=5000$ even if the precision $\hat\epsilon_f$ has not been reached yet.
\item The closed inversion formula \eqref{mplestimation} is applied, as if the coupling potential was not present:
\be
z_k^{**} \,=\,\frac{\avg{\alpha_k}_{z,J}} {\avg{\prod_{v\in k}\alpha_v}_{z,J}}\,,\quad k\in K \;.
\ee
\item We study the parameter
\be
 \delta \,=\, \frac{1}{|K|}\sum_{k\in K} \big( \log z_k^{**} - \log z_k^{*} \big) \;.
\ee
For zero coupling potential $\delta$ is close to zero, since both $z_k^{**}$ and $z_k^*$ equal the true value of the activity $z_k$ (up to the precision of the gradient descent algorithm).
\end{itemize}

\begin{figure}
  \centering
  \includegraphics[width = 0.85\textwidth]{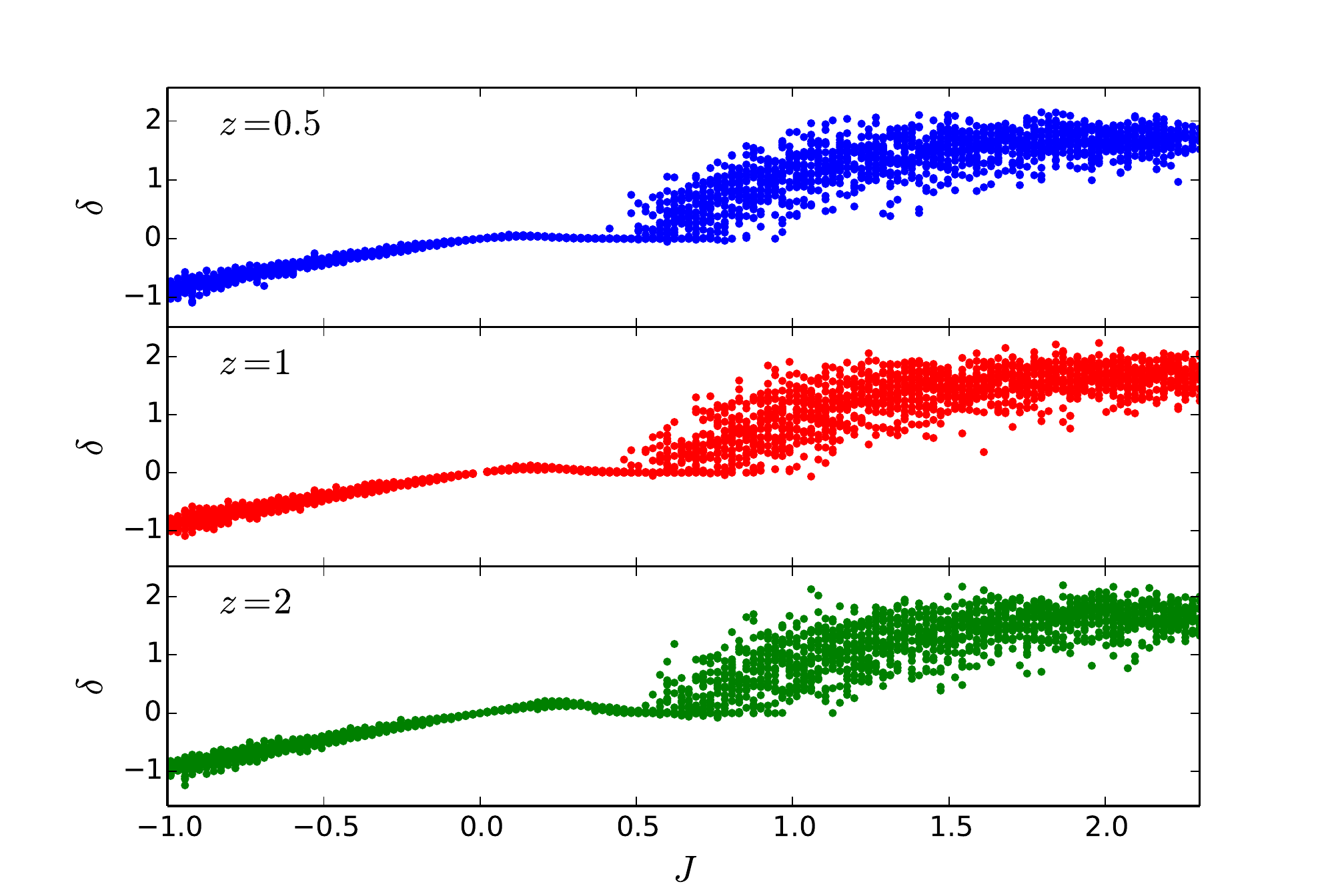}
  \includegraphics[width = 0.85\textwidth]{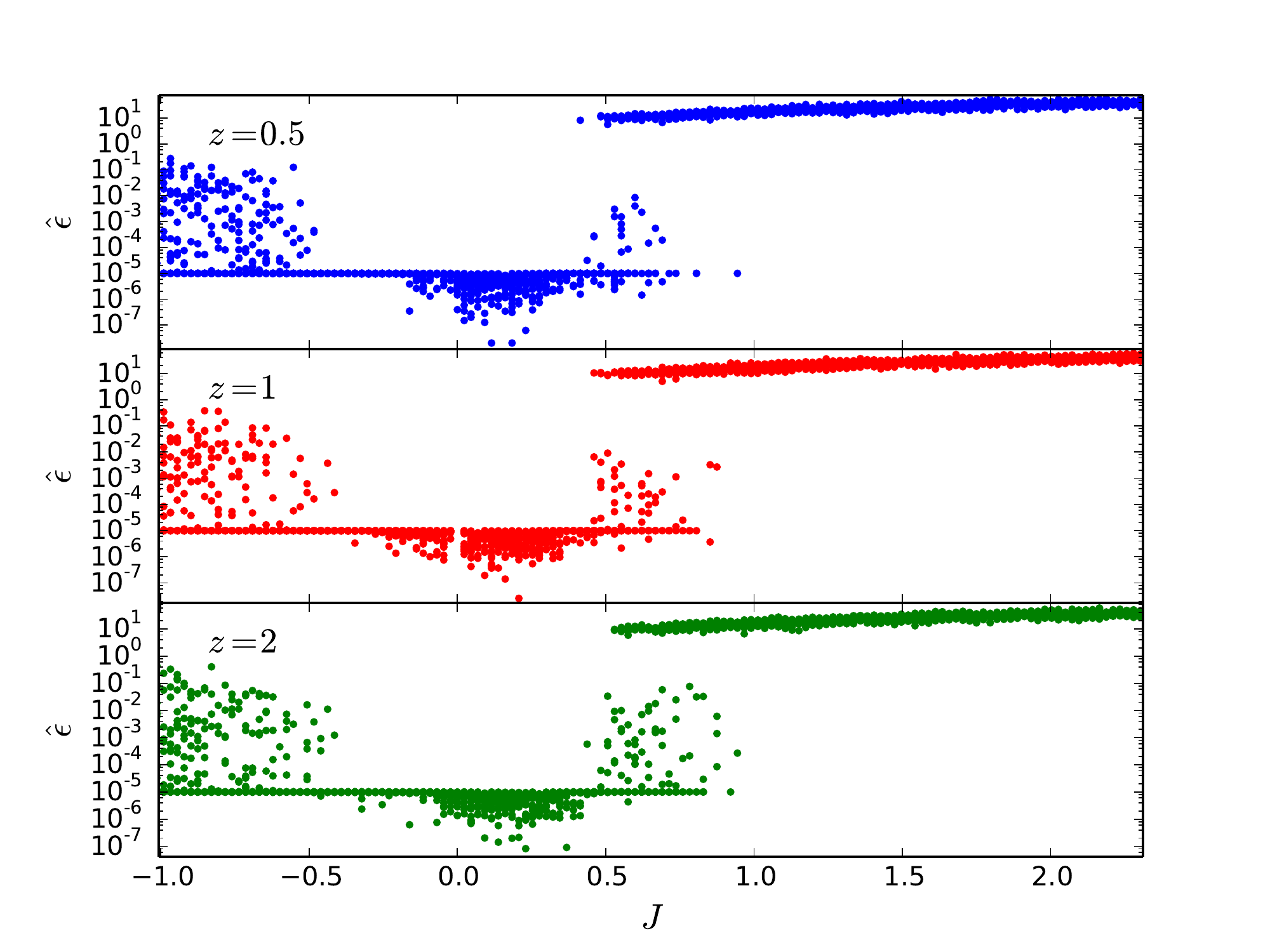}
  \caption{\textbf{Tests for the presence of imitative interaction.} (Colour online)
\textit{On top:} Parameter $\delta=\frac{1}{|K|}\sum_{k\in K} (\log z_k^{**} - \log z_k^*)$ evaluated through the use of both the analytic inversion formula and the gradient descent method, as if the imitative interaction was not present, versus imitative potential $J$. \textit{A value $\delta<0$ reveals that $J<0$. On the other hand, the order of magnitude of $\delta$ and its variance grow abruptly when $J$ crosses a positive critical value.}
\textit{On bottom:} Precision $\hat\epsilon = \max_{k\in K}|\log\avg{\alpha_k}_{z^*}-\log\avg{\alpha_k}_{z,J}|$ of the gradient descent algorithm built as if the imitative interaction was not present, versus imitative potential $J$ (linear-log scale). \textit{The convergence is always reached for $J$ close to $0$, while it is never reached for $J$ larger than a critical value.}
The polymer activity and the imitative potential are the same for each hyper-edge: $z_k=z,\; J_k=J\; \forall\,k\in K$. For each value of $J$ we performed $20$ trials on different random hypergraphs, taking always $N = 20$, $c_1 = c_2 = 1$ and $z=0.5$ (blue), $z=1$ (red), $z=2$ (green).}
  \label{fig:J delta}
\end{figure}

We observed that $\delta$, together with the final precision $\hat\epsilon$, can indeed be used as a test-parameter to understand whether the real system obeys a pure hard-core interaction or there are other types of non-negligible interactions. In fact it allows to distinguish between the following three regimes (Fig. \ref{fig:J delta}):
\begin{itemize}
\item For $J<0$ the gradient descent algorithm is not guaranteed to converge in the prescribed number of iterations since the precision $\hat\epsilon$ ranges from $10^{-5}$ to $10^0$. The value of $\delta$ is negative and its modulus grows linearly with $J$.
\item For $0<J<J_0$ the convergence of the gradient descent method is attained. The parameter $\delta$ is close to zero, positive, and shows a non-monotonic behaviour in $J$.
\item For $J>J_0$ the convergence of the gradient descent method becomes abruptly poor and for $J$ sufficiently large $\hat\epsilon$ is larger that $10^1$. $\delta$ is positive and exhibits a large variance over different random hypergraphs.
\end{itemize}

\begin{figure}
  \centering
  \includegraphics[width = 0.85\textwidth]{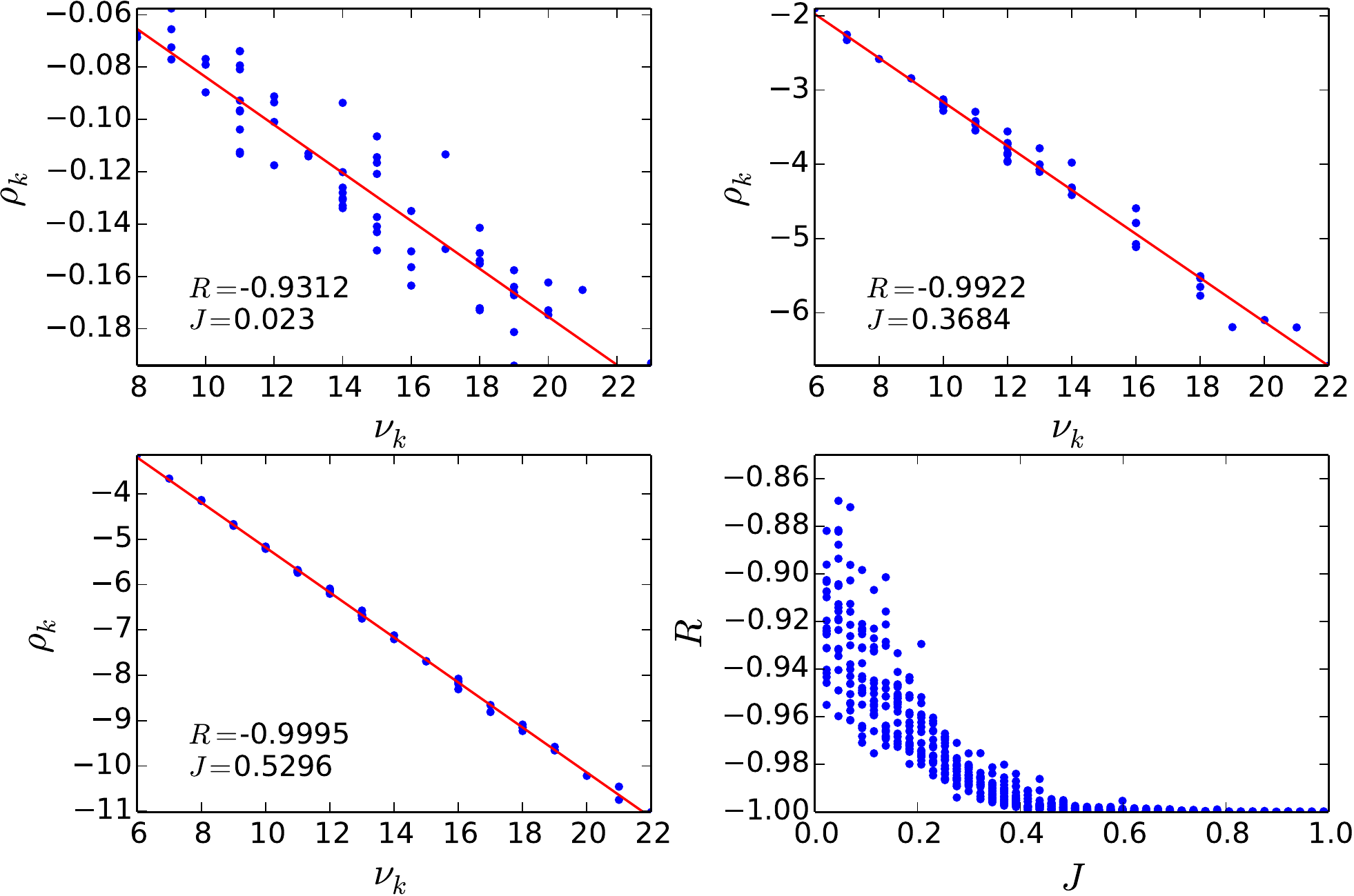}
  \includegraphics[width = 0.85\textwidth]{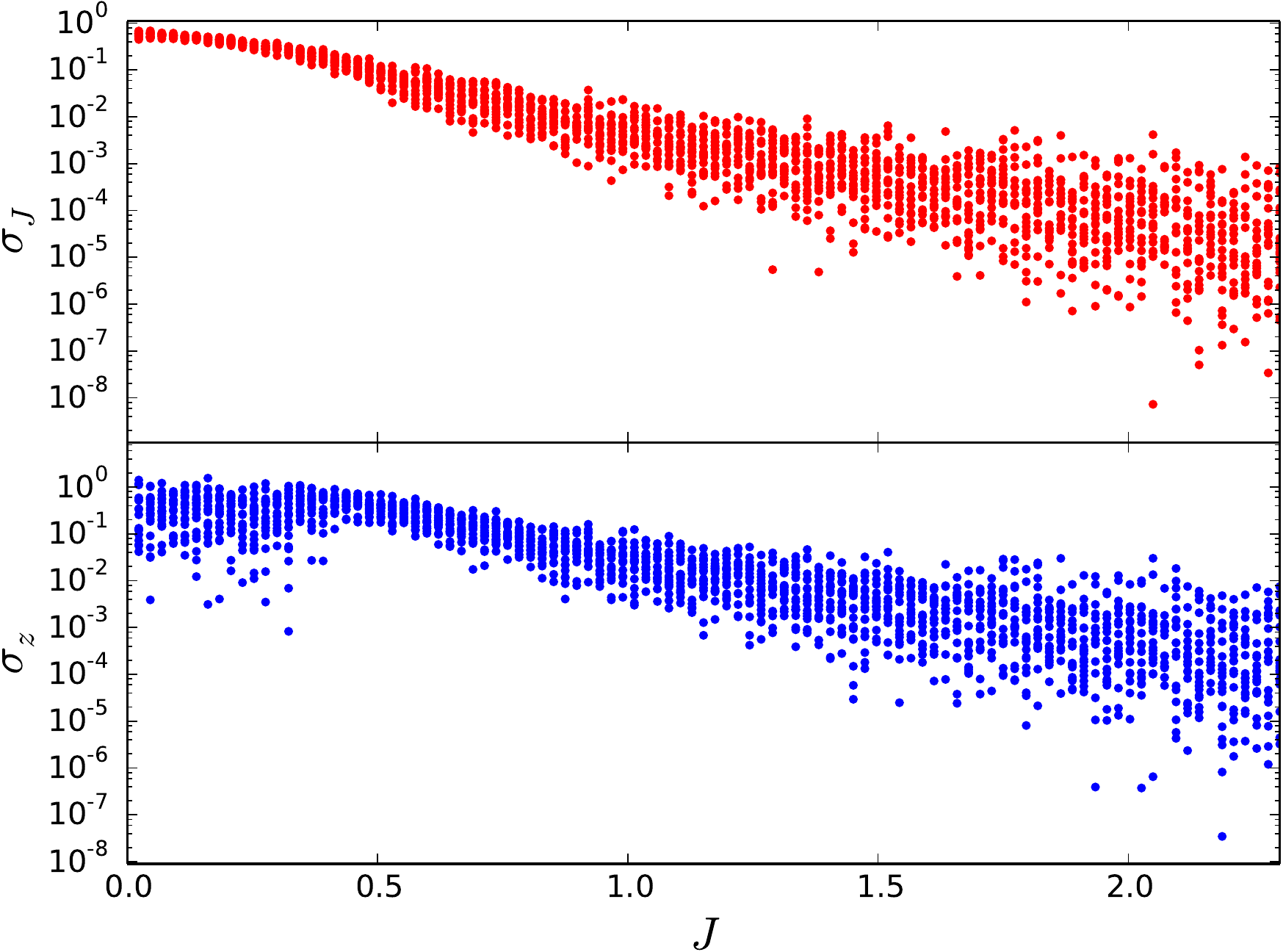}
  \caption{\textbf{Estimate of the imitative potential $J$.} (Colour online)
\textit{On top:} parameter $\rho_k=\log \avg{\alpha_k}/\avg{\prod_{v\in k}\alpha_v}$ versus $\nu_k=\textrm{Card}\{h\in K\,|\,|h\cap k|>0\}$ for every hyperedge $k$ in a random hypergraph (blue dots).
The polymer activity and the coupling are the same for each hyperedge: $z_k=z,\; J_k=J\; \forall\,k\in K$. The test is performed on a random hypergraph taking $N = 25$, $c_1 = c_2 = 1$, $z=1$ and different values of $J$: $J=0.023$, $J=0.3684$, $J=0.5296$. \textit{The relation between $\rho_k$ and $\nu_k$ is linear for $J$ sufficiently large}: a linear fit (red line) is performed according to the relation $\rho_k = -\alpha\,\nu_k+ \beta$. The reliability of this fit is tested by plotting the correlation coefficient $R$ versus $J$.
\textit{On bottom:} relative errors $\sigma_J=\left|\frac{\alpha-J}{J}\right|$ (red) and $\sigma_z=\left|\frac{\beta-\log z}{\log z}\right|$ (blue), versus $J$. \textit{According to the relation \eqref{eq:stima2}, the slope of the fit $\alpha$ is used as an estimate of the coupling $J$, when $J$ is sufficiently large.}}
  \label{fig:J fit}
\end{figure}

When $J$ is positive and sufficiently large, we propose a method to estimate its value.
Compare the relations \eqref{eq:stima} for the measure $\mu_{z,J}$ with the exact relations \eqref{hcrelations} for the measure $\mu_z$. It becomes clear that if the experimental parameter $\rho_k \equiv \log\big(\avg{\alpha_k}_{\textrm{exp}}/\avg{\prod_{v\in k}\alpha_v}_{\textrm{exp}}\big)$ shows a correlation with the number of hyperedges intersecting $k$, $\nu_k\equiv\textrm{Card}\{h\in K,\,|h\cap k|>0\}$, then the system presents other interactions beyond the hard-core one.
In particular in the case of constant $J$ and $z$, the equation \eqref{eq:stima} gives
\be \label{eq:stima2}
 \rho_k(z,J) \,\approx\, \log z - J\,\nu_k \,, \quad k\in K
\ee
when $J\,\textrm{Card}\{q\in K,\,q\ni v,\,|q\cap k|=0\}$ is sufficiently large with respect to $\log z_p$, for all hyperedges $p$ intersecting $k$ and all vertices $v$ neighbouring $k$.
Therefore $J$ and $z$ can be found by performing a linear fit between $\rho_k$ and $\nu_k$ (Fig. \ref{fig:J fit}).\\

\section{Conclusions and Outlooks}

With the purpose to investigate the possibility to discover topological information from the data space we introduced
in this work a model in which polymers are deposited on the hyperedges of an hypergraph with a probability determined according to the hyperedges activities. The idea underlying the model is that simple graphs are no longer able to account for the structure of many modern socio-technical systems, such as those of virtual messaging systems or voip calls. In these systems the communications do not occur only between pairs of users, but may involve larger groups \cite{GinestraSimplex}. We believe that this context may give rise to new interesting behaviours, where topology plays a crucial role.

With these applications in mind we tackled the inverse problem. After finding an extension of the Heilmann-Lieb relations 
that fits the higher-dimensional case, we introduced the \textit{Maximum Likelihood Estimation} (MLE) and the \textit{Maximum Pseudo-Likelihood Estimation} (MPLE) solutions of the inverse problem. While the latter constitutes a more rough estimate but has an explicit form in terms of experimental quantities, the former provides a more precise but implicit solution, which can nonetheless be numerically evaluated by the gradient descent algorithm we proposed. We found that by introducing a variable update step size the algorithm converges with exponential precision in the number of steps. However we stress that the time it takes to compute each step of the algorithm grows with the size of the hypergraph, since all the admissible configurations have to be computed exactly. A possible solution to this problem could be to evaluate average quantities through Markov chain Monte Carlo sampling. We tested the algorithm on toy models for different values of the parameters, and found that while the exponential convergence does not seem to be influenced by the number of vertices in the hypergraphs, it does depend on the values of the activities. A further analysis of this dependence could be performed, for example with respect to the variance of the activity distribution.

We then considered the presence of an interaction between the monomers in the configurations. The meaning of this interactions can be understood by thinking to the social systems that our model tries to describe: in the context of virtual social interactions peer-to-peer effects are to be expected. We found that a comparison between the MLE and the MPLE solution of the inverse problem can be used to detect the presence of such an interaction. The same comparison can moreover lead to the estimation of the interaction magnitude in the ``strong interaction'' regime.

The next step and most natural continuation of this work would be the application of such a model on real-world data. By testing the model on data we could verify whether it is able to accurately describe the behaviour of users in virtual messaging services
and what type of predictive ability it comes with. For instance, this could be done by measuring the Kullback-Leibler distance between the experimental probability distribution and the probability distribution resulting from the Maximum Likelihood Estimation. In case the model is accurate it would allow us to measure of user activities in chatrooms, and even determine whether the system is subject to peer-to-peer monomer interactions.

\vskip 0.5cm

{\bf Aknowledgments} The authors are deeply indebted to Mario Rasetti for inspiring this work
and for many illuminating discussions. We also thank Massimo Ferri, Giovanni Petri,
Federico Ricci-Tersenghi, Alina S\^irbu and Francesco Vaccarino for interesting discussions.
This work was partially supported by FIRB (grant number RBFR10N90W), PRIN (grant number 2010HXAW77) and INdAM-GNFM (Progetto Giovani 2015).

\appendix
\section*{Appendix}

We shortly present here the application of the Maximum Likelihood and the Maximum Pseudo-Likelihood Methods to our model. The general framework is the following: fix the hypergraph $H$ and assume the model is described by an unknown value of the activities $z$ to be determined.
Consider a set of $M$ observations of polymer configurations $\bar{\alpha}=\{\alpha^{(s)}\}_ {s=1,\ldots, M}$, where   $\alpha^{(s)}=(\alpha^{(s)}_k )_{k\in K}$ and $\alpha^{(s)}_k$ encodes the presence/absence of a polymer on the hyperedge $k$ in the $s^{\mathrm{th}}$ experimental observation.
Suppose that  $\bar{\alpha}$ is a set of independent observations sampled from the same probability distribution $\mu_{z}$,
for a certain value of the activities $z=z^*$.

We use two standard methods that give an optimal value $z^*$ to fit the dataset $\bar{\alpha}$: the \textit{maximum likelihood estimation} (MLE) and the \textit{maximum pseudo-likelihood estimation} (MPLE). Let us briefly recall these methods.

The optimal estimate $z^*$ in the MLE sense maximizes the \textit{likelihood function} defined as
\begin{equation}\label{AMLE}
	\mathcal{L}(z;\bar{\alpha}) \,=\, \prod_{s = 1}^{M} \mu_{z}(\alpha^{(s)}) \;.
\end{equation}
Standard computations show that $\log \mathcal{L}(z;\bar{\alpha})$ is a concave function in the variables $\log z$ and it attains its maximum at the point $z^*$ satisfying the following system of $|K|$ equations:
\be\label{Amlestimation}
\dl \alpha_k\dr_{z^*}=\dl \alpha_k \dr_{\mathrm{exp}}\,,\quad k\in K=E\cup F \;,
\ee
where $\dl \alpha_k \dr_{\mathrm{exp}} \equiv \frac{1}{M}\sum_{s=1}^M \alpha^{(s)}_k$ is the experimental average value of the presence of a polymer in the hyperedge $k$. This approach naturally fits the experimental situation where the available data is the set of \textit{empirical polymer densities}.
Let us observe that the likelihood function $\mathcal{L}(z;\bar\alpha)$ is strictly related to the Kullback-Leibler divergence of the measure $\mu_z$ from the empirical measure $\mu^*$, defined as
\be
D_{\mathrm{KL}}(\mu_z|\mu^*) \,=\, \sum_{\alpha} \mu^*(\alpha)\,\log\frac{\mu^*(\alpha)}{\mu_z(\alpha)}
\ee
where $\mu^*(\alpha)\equiv\frac{1}{M}\sum_{s=1}^M\delta(\alpha=\alpha^{(s)})\,$. Precisely the following relations holds:
\be
\frac{1}{M}\log\mathcal{L}(z;\bar\alpha) \,=\, - D_{\mathrm{KL}}(\mu_z|\mu^*) \,+\, C
\ee
with $C=\sum_{\alpha}\mu^*(\alpha)\log\mu^*(\alpha)\,$.

Now let us consider the pseudo-likelihood instead of the likelihood. The optimal estimate $z^*$ in the MPLE sense maximizes the \textit{pseudo-likelihood function} defined as
\begin{equation}\label{AMPLE}
	\mathcal{L}^P(z;\bar{\alpha}) \,=\,  \prod_{s = 1}^{M} \prod_{k \in K} \mu_{z}\big(\alpha_k^{(s)}\big|\alpha_{\neq k}^{(s)}\big)
\end{equation}
where, for a given sample $s$ and hyperedge $k$, $\alpha_{\neq k}^{(s)}$ encodes the experimental observation of a polymer on all the hyperedges different from $k$.
It is possible to show that $\mathcal{L}^P$ attains its maximum at the point $z^{**}$ explicitly defined by the following $|K|$ conditions:
\be\label{Amplestimation}
\dl \alpha_k \dr_{\mathrm{exp}} \,=\, z^{**}_k\, \big\dl \prod_{v\in k} \alpha_v \big\dr_{\mathrm{exp}}\,,\quad k\in K=E\cup F \;
\ee
where $\alpha_v^{(s)}$ denotes the experimental observations of a monomer on the vertex $v$ in the $s^{\textrm{th}}$ trial and $\dl \prod_{v\in k}\alpha_v \dr_{\mathrm{exp}} \equiv \frac{1}{M}\sum_{s=1}^M \prod_{v\in k}\alpha^{(s)}_v$ is the empirical monomer correlation of the vertices in $k$.

\end{document}